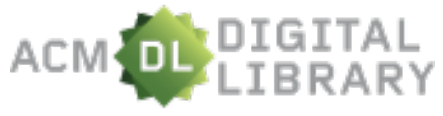
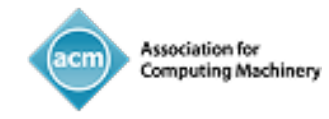
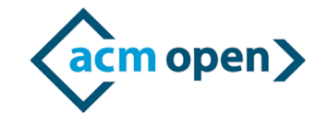

RESEARCH-ARTICLE

# C2-Cite: Contextual-Aware Citation Generation for Attributed Large Language Models

**YUE YU**, Beijing University of Posts and Telecommunications, Beijing, Beijing, China

**TING BAI**, Beijing University of Posts and Telecommunications, Beijing, Beijing, China

**HENGZHI LAN**, Beijing University of Posts and Telecommunications, Beijing, Beijing, China

**LI QIAN**

**LI PENG**

**JIE WU**

**View all**

# $C^2$-Cite: Contextual-Aware Citation Generation for Attributed Large Language Models

Yue Yu
Beijing University of Posts and Telecommunication
Beijing, China
loadingyy@bupt.edu.cn

Ting Bai*
Beijing University of Posts and Telecommunication
Beijing, China
baiting@bupt.edu.cn

Hengzhi Lan
Beijing University of Posts and Telecommunication
Beijing, China
lansnowz@bupt.edu.cn

Li Qian
Researcher
Beijing, China
qianli_xfjj@163.com

Li Peng
Researcher
Beijing, China
qrsssh@163.com

Jie Wu
Researcher
Beijing, China
jiewu2017@gmail.com

Wei Liu
Researcher
Beijing, China
buptliuwei@gmail.com

Jian Luan
Researcher
Beijing, China
jian.luan@gmail.com

Chuan Shi*
Beijing University of Posts and Telecommunication
Beijing, China
shichuan@bupt.edu.cn

## Abstract

The attribution technique enhances the credibility of LLMs by adding citations to the generated sentences, enabling users to trace back to the original sources and verify the reliability of the output. However, existing instruction-tuned attributed LLMs often fail to properly interpret the contextual semantics of citation symbols (e.g., [i]) during text generation. This shortcoming arises from their insufficient awareness of the context information surrounding citation markers, which in turn leads to disjointed references and poor integration of retrieved knowledge into the generated content. To address this issue, we propose a novel **C**ontextual-aware **C**itation generation framework (**$C^2$-Cite**) that explicitly integrates the semantic relationships between citation markers and their referenced content. Specifically, a contextual citation alignment mechanism is adopted: it first encodes the retrieved document contexts into the symbol representation of citations, then aligns the marker numbers by decoding information from a citation router function. This mechanism enables the transformation of citation markers from generic placeholders into active knowledge pointers that link to the referenced source information. Experimental results on the ALCE benchmark across three datasets validate our framework $C^2$-Cite++: it outperforms the SOTA baseline by an average of 5.8% in citation quality and 17.4% in response correctness. The implementation is publicly available at https://github.com/BAI-LAB/c2cite

*Corresponding author.

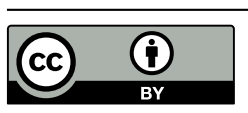

## 1 Introduction

Recent advances in Retrieval-Augmented Generation (RAG) have significantly reduced hallucinations in Large Language Models (LLMs) by effectively incorporating external knowledge sources [2–4, 9, 12, 18, 21, 39]. However, ensuring the long-term reliability and trustworthiness of the generated content remains a persistent and particularly challenging issue in the field. Considerable research efforts have been dedicated to integrating attribution techniques into large language models (LLMs) to further bolster the credibility of their generated content. These techniques primarily involve appending specific citations to model outputs—explicitly linking the generated content of LLMs to relevant external information sources (such as academic documents, datasets, or authoritative databases) as concrete supporting evidence.

Existing attributed LLMs can be generally classified into three distinct types: in-context learning attribution [19, 23, 37, 40, 44], post-hoc retrieval [7, 25, 28–30, 42], and instruction-tuned attribution [15, 16, 31]. The first two approaches, however, lack inherent attribution capabilities without extensive fine-tuning and often fail to dynamically align citation information during the complex generation process, leading to disjointed citations and low-quality references that undermine the overall reliability [38].

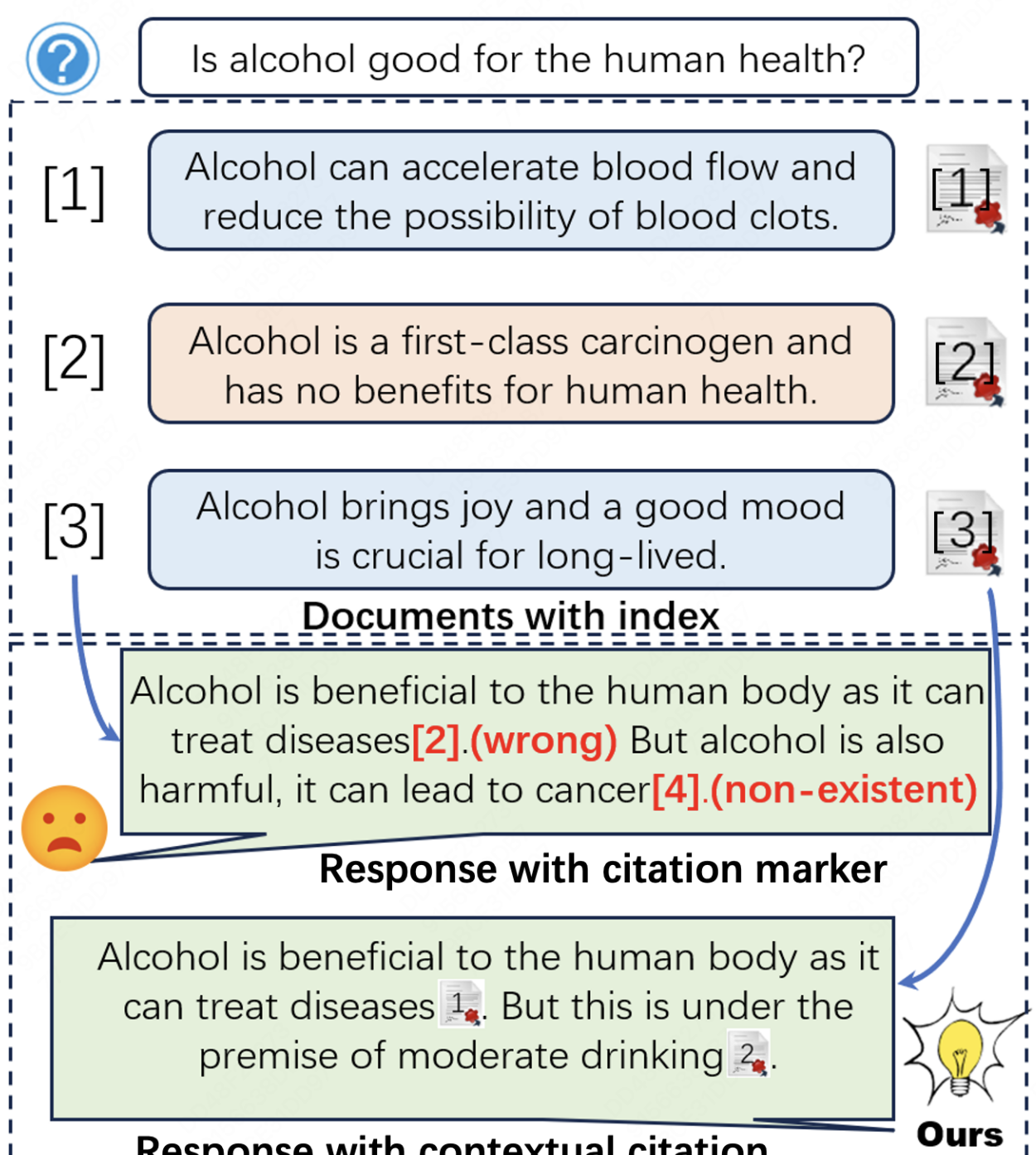

**Figure 1: Illustration of integrating contextual citation (i.e., ours) v.s. using only citation markers (i.e., [i]) in response generation of LLMs. Without contextual information, LLMs may produce erroneous or fictitious citations.**

Recent advances in instruction-tuned attributed models [15, 16, 41, 43] have primarily focused on developing comprehensive pipelines for constructing high-quality fine-tuning datasets, positing that well-curated instruction-tuning data can enable LLMs to perform accurate factual attribution. However, these methods rely exclusively on the inherently weak and unstable associations between the semantics of citation symbols (e.g., [i]) and their referenced content. This over-reliance on such fragile connections frequently leads to misinterpreted citation meanings or the generation of non-existent symbols (as illustrated in Fig. 1), which severely compromises the overall citation quality. This critical issue likely stems from the nature of LLM pre-training, where such symbols are typically treated as generic placeholders rather than context-sensitive pointers that directly reference specific external knowledge.

To address these issues, we propose a novel context-aware citation generation framework **C$^2$-Cite** for attributed LLMs. We introduce a context-aware citation alignment mechanism to explicitly integrate the semantic relationships between citation markers and their referenced content. Specifically, C$^2$-Cite innovates by encoding citation symbols with contextual information from retrieved documents during fine-tuning, enabling LLMs to associate citation symbols (e.g., [i]) with specific semantic content. Additionally, a citation token classifier identifies citation positions and aligns the symbol with its referenced text, ensuring that citation symbols (e.g., [i]) function as active knowledge pointers. Our work emphasizes the contextual alignment of citations during the generation process and significantly improves the quality of attributed LLMs by enhancing their contextual awareness. The contributions are summarized as follows:

- We propose C$^2$-Cite, a novel fine-tuning attributed LLM framework that explicitly models the semantic relationships between citation symbols and their referenced contextual content.
- We innovatively integrate contextual-aware embedding and contextual citation alignment components to transform citation symbols from passive placeholders into active knowledge pointers, significantly enhancing response and citation quality.
- Experimental results on the ALCE benchmark across three datasets validate the effectiveness of our framework C$^2$-Cite++: it achieves average improvements of 5.8% and 17.4% over the SOTA baseline Front in citation quality (F1 score) and response correctness metrics.

## 2 Related Work

### 2.1 In-Context Learning Attribution

The foundational approach to in-context learning attribution is Prompt-based In-Context Learning models (P-ICL), where attribution is accomplished via structured prompt engineering rather than through iterative refinement. These methods construct input prompts integrating queries and source documents, instructing models to generate responses with explicit citations, like appending document labels to each sentence[6, 17, 26, 27]. For example, the ALCE [11] benchmark serves as a canonical P-ICL baseline, establishing a standardized input format for attribution tasks by embedding citation requirements directly into prompt templates. Besides, certain efforts also employ iterative processes to generate responses, and they try to rewrite the answer through iteration. The approach in [34] iteratively generates both sentences and references, whereas the method proposed in [36] leverages a repeated retrieval process to refine document selection, thereby enhancing response quality and attribution accuracy in multi-agent systems. These methods are not only time-consuming due to their iterative nature but also heavily dependent on the model's inherent capabilities, resulting in suboptimal performance.

### 2.2 Instruction-Tuning Attribution

Instruction-tuning attribution approaches [5, 7, 28, 33] focus on generating high-quality training datasets to fine-tune LLM models for improved attribution performance. These methods operate under the assumption that better attributed training data yields models with stronger attribution abilities. High-quality citation-annotated sentences are generated using LLMs, followed by filtering out low-quality data instances according to predefined rules to construct fine-tuning datasets [31]. For example, Front [16] utilizes GPT-generated citation-annotated responses, where the outputs serve as positive samples, while responses from a weaker language model act as negative samples for reinforcement learning via Direct

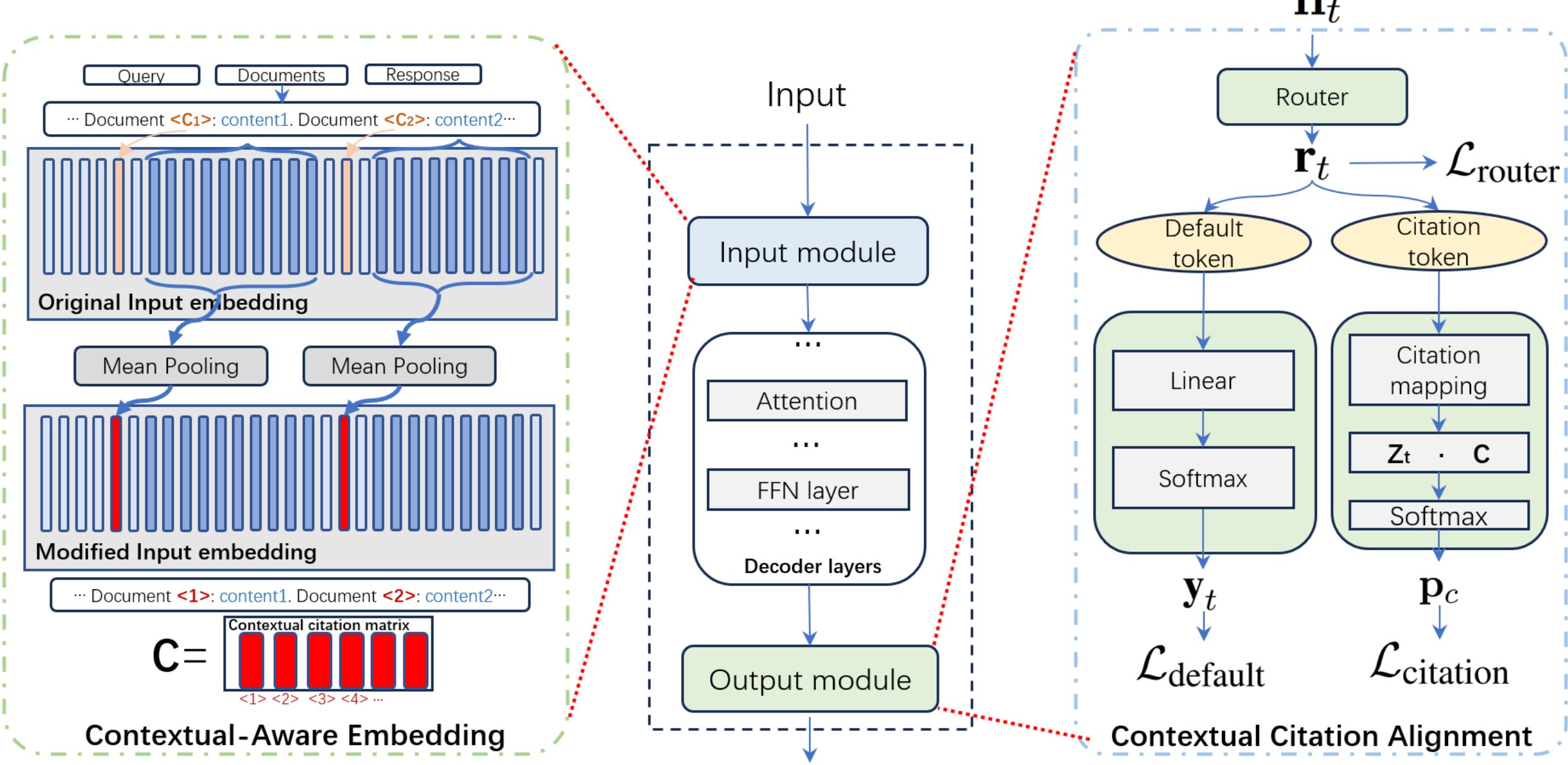

**Figure 2: The architecture overview of C[2]-Cite with two key components: Contextual-Aware Embedding and Contextual Citation Alignment components to transform passive placeholders into active knowledge pointers for citation symbols.**

Preference Optimization (DPO) training. In a similar pipeline, fine-grained rewards are used in [15] to refine GPT-generated citation-annotated outputs via Reject Sampling (RS) to select high-quality data for better model performance.

These methods heavily rely on the capabilities of strong LLM models or well-designed workflows to generate high-quality training data. Different from them, our framework C[2]-Cite emphasizes designing effective citation learning methods that can accurately integrate citation sources into response generation, with minimal reliance on complex preprocessing pipelines.

## 2.3 In-Generation vs. Post-Hoc Attribution

Based on when citations are integrated, two primary approaches exist for generating responses with citations: simultaneous citation-inclusive generation and post-hoc retrieval attribution. The former uses either prompt engineering (e.g., P-ICL methods) or instruction tuning to guide models to append citation labels during content generation, achieving efficient and straightforward single-pass attribution through the model's native prompting abilities, yet it is reliant on the model's precision in parsing prompt formats [10, 20, 22, 24, 34, 39]. The latter first generates citation-free responses, then retroactively traces sources by analyzing model-internal signals such as hidden states or attention patterns. This approach excels in fine-grained interpretability and adapts to untuned models but incurs computational costs due to its two-stage processing [7, 28]. Functioning more as interpreters than intrinsic attribution mechanisms, these attribution methods focus more on post-hoc analytical depth.

Our framework C[2]-Cite is an instruction-tuning attribution method within the in-generation paradigm. This design retains the efficiency of single-pass generation while strengthening the intrinsic semantic connection between citation markers and their referenced content, overcoming over-reliance on prompt format parsing.

# 3 Methodology

## 3.1 Problem Definition

Attributed Large Language Models (Attributed LLMs) refer to the models that produce content interspersed with explicit citation markers (e.g., [i]), which are directly linked to specific external information sources (e.g., documents) in a retrieval corpus. Given a query $Q$ and a pre-retrieved document corpus annotated with citation numbers $\mathcal{D} = \{d_{[1]}, d_{[2]}, ..., d_{[N]}\}$, the aim of attributed LLMs is to generate a response to the query that contains sentences with corresponding citations from the document corpus $\mathcal{D}$, denoted as:

$$\mathcal{R} = \{\ldots, s_0, c_0, s_1, c_1, \ldots\}, \tag{1}$$

where $s_i$ denotes a sentence and $c_i$ represents the citation marker (e.g., $[i]$) referring to document $d_{[i]} \in \mathcal{D}$.

An attributed LLM is capable of effectively leveraging information within the retrieval corpus to generate well-grounded and contextually aligned responses.

## 3.2 The Overview of C[2]-Cite

The overview architecture of C[2]-Cite is shown in Fig. 2. Our framework is constructed upon a standard decoder-only LLM architecture that comprises a stack of Transformer decoder layers. To achieve contextual-aware citation embeddings, we modify the input and output modules. Specifically, C$^2$-Cite incorporates two key components: a **Contextual-Aware Embedding** component (incorporating contextual information into citation markers for semantic relevance) and a **Contextual Citation Alignment** component (optimizing fine-tuning to align generated text with retrieval documents) to transform passive placeholders into active knowledge pointers for citation symbols.

## 3.3 Contextual-Aware Embedding

To enhance the model's ability to leverage source information, we introduce a contextual embedding fusion mechanism to model contextual text information into citation markers, infusing them with semantic relevance to the surrounding content. Specifically, we normalize citation markers by converting the multi-token form (i.e., "[i]") into single special tokens (i.e., "$\langle c_i \rangle$"), where $\langle c_i \rangle$ refers to the citation marker of document $d_{[i]}$. Unlike default LLMs that treat citation tokens as placeholders without external semantic meaning, we explicitly assign *contextual embeddings* (i.e., embeddings that encode both the semantic information of the surrounding textual context and the positional relevance to the corresponding retrieved document in the retrieval corpus) to each citation marker $\langle c_i \rangle$. As shown in Fig. 2, for citation marker token $\langle c_i \rangle$, we compute the contextual embedding from the corresponding document $d_{[i]}$ to replace its symbolic embedding, formulated as:

$$\mathbf{c}_i = \mathcal{F}_{contextual}(\mathbf{e}_1^{(i)}, \mathbf{e}_2^{(i)}, \dots, \mathbf{e}_N^{(i)}), \tag{2}$$

where $\mathbf{c}_i$ is the contextual embedding of citation marker $\langle c_i \rangle$, $\mathbf{e}_j^{(i)} \in \mathbb{R}^{1\times H}$ represents the embedding of $j$th token in $d_{[i]}$, $N$ is the number of tokens. Mean-pooling operation is used as the aggregation function $\mathcal{F}_{contextual}$.

By replacing citation symbols with document semantic information (via mean-pooling), our approach allows the model to access the semantic information of the cited document directly through the citation token, enhancing source grounding during generation. This operation mitigates potential citation errors arising from pretrained LLMs treating such markers as generic placeholders, as the explicit integration of document semantics ensures more accurate contextual interpretation.

## 3.4 Contextual Citation Alignment

To enable LLMs to learn attribution, we introduce a contextual citation alignment mechanism that integrates contextual citation markers into response generation. During fine-tuning, it optimizes text generation to enable accurate association between responses and retrieval documents. Specifically, given the hidden state $\mathbf{h}_t \in \mathbb{R}^{1\times H}$ of token $t$ from the decoder layer, a citation router function determines whether the token is a default token or a citation token. The router function is implemented as a binary classifier, defined as:

$$\hat{\mathbf{r}}_t = \text{Softmax}(\mathbf{W}_r \cdot \mathbf{h}_t^\top), \tag{3}$$

where $\mathbf{W}_r \in \mathbb{R}^{2\times H}$ is the trainable parameter matrix, and $\hat{\mathbf{r}}_t$ represents the probabilities for default or citation token.

The citation router is optimized by a binary cross-entropy loss function, formulated as:

$$\mathcal{L}_{\text{router}} = \mathcal{F}_{BCE}(r_t, \hat{r}_t), \tag{4}$$

where $\mathcal{F}_{BCE}$ is the binary cross-entropy loss function. The true label of token $t$ is determined by:

$$r_t = \begin{cases} 0, & \text{if } r_t \in \mathcal{T}_{\text{default}} \\ 1, & \text{if } r_t \in \mathcal{T}_{\text{citation}}, \end{cases} \tag{5}$$

where $\mathcal{T}_{\text{default}}$ and $\mathcal{T}_{\text{citation}}$ represent the set of default and citation tokens, respectively.

*3.4.1 The optimization of default tokens.* Consistent with optimizations in the general LLMs generation processes, the output layer uses a linear layer to project the decoded hidden states into vocabulary logits, followed by a softmax to convert these into token probability distributions. During training, cross-entropy loss measures the probabilities between the predicted token and the true token, driving model predictions to align with target outputs. The loss function is defined as $\mathcal{L}_{\text{default}}$.

*3.4.2 The optimization of citation tokens.* The attribution capability of LLMs is incorporated by aligning the contextual-aware representation of citation markers with their true citation symbolic numbers. Specifically, given the hidden state vector $\mathbf{h}_t$ from the decoder layer of a citation token, we first map it into the citation space, defined as:

$$\mathbf{z}_t = \mathbf{W}_c \cdot \mathbf{h}_t^\top + b_c, \tag{6}$$

where $\mathbf{W}_c \in \mathbb{R}^{H\times H}$ is the transformation matrix, and $\mathbf{z}_t \in \mathbb{R}^{H\times 1}$ is the transferred citation representation.

Then we compare all contextual-aware citation embeddings $\{\mathbf{c}_1, \mathbf{c}_2, ..., \mathbf{c}_N\}$ of citation markers $\{\langle c_1 \rangle, \dots, \langle c_N \rangle\}$ (defined in Eq. 2) with embedding $\mathbf{z}_t \in \mathbb{R}^{H\times 1}$ of the current citation token. Our aim is to classify the citation token $\langle c_t \rangle$ into the true citation marker number according to their semantic similarity, defined as:

$$\mathbf{p}_c = \text{Softmax}(\mathbf{C} \cdot \mathbf{z}_t), \tag{7}$$

where $\mathbf{C} = [\mathbf{c}_1, \mathbf{c}_2, ..., \mathbf{c}_N] \in \mathbb{R}^{N\times H}$ is the contextual-aware citation matrix, and $\mathbf{p}_c \in \mathbb{R}^{N\times 1}$ represents the probability of each citation marker that the current citation token belongs to.

The loss function of citation alignment can be defined as:

$$\mathcal{L}_{\text{citation}} = \mathcal{F}_{CE}(p_c, \hat{p}_c), \tag{8}$$

where $\mathcal{F}_{CE}$ is the cross-entropy loss function, $p_c$ is the true citation marker number.

## 3.5 Contextual Attentive Augment

In our approach, citations evolve from passive symbolic markers into semantically aware elements that actively guide content generation. To enhance the utilization of such contextual-aware citation information, we introduce a contextual attentive augmentation mechanism to amplify the model's attention on previously generated citation tokens. This ensures that our model better integrate the contextual information of citations when generating subsequent tokens.

Specifically, given a response fragment composed of tokens, e.g., $\mathcal{R} = \{\dots, t_c, \dots, t_j \dots\}$. For subsequent default tokens $t_j$, the attention score $S_{(t_j, t_c)}$ between $t_j$ and the citation token $t_c$ is constrained to approximate the target attention score $\hat{S}(t_j, t_c)$. The attention score in default LLMs is defined as:

$$S(t_j, t_c) = \frac{Q_j \cdot K_c}{|Q_j| \cdot |K_c|}, \quad (9)$$

where $Q$ and $K$ are the query and key of a token, respectively. The citation-aware attention score is defined as:

$$\hat{S}(t_j, t_c) = \lambda_{j-c}^{distance} \cdot A(t_j, t_c), \quad (10)$$

where $\lambda_{j-c}^{distance} \propto \frac{1}{j-c}$ represents the distance-aware decay coefficient, which uses inverse distance weighting to reduce attention allocation to distant citation tokens. The empirical setting of the total attention score $A(t_c, t_j)$ is 0.3, and the allocation is performed proportionally to the positional distance $|j - c|$.

The loss function to augment the contextual citation attention is optimized by:

$$\mathcal{L}_{\text{attn}} = \frac{1}{|\mathcal{T}_{\text{default}}| \cdot |\mathcal{T}_{\text{citation}}|} \Sigma_{t_j, t_c \in \mathcal{R}} |\hat{S}_{(t_j, t_c)} - S_{(t_j, t_c)}|, \quad (11)$$

where $|\mathcal{T}_{\text{default}}|$ and $|\mathcal{T}_{\text{citation}}|$ represent the total number of default and citation tokens, respectively.

## 3.6 Model Optimization

The final loss of C$^2$-Cite is a weighted sum of the above default loss in LLMs (i.e., $\mathcal{L}_{\text{default}}$), citation alignment loss (i.e., $\mathcal{L}_{\text{citation}}$), router loss (i.e., $\mathcal{L}_{\text{router}}$), and citation attentive loss (i.e., $\mathcal{L}_{\text{attn}}$), ensuring attributed response generation with contextual-aware citation information integration, defined as:

$$\mathcal{L} = \mathcal{L}_{\text{default}} + \alpha\mathcal{L}_{\text{citation}} + \beta\mathcal{L}_{\text{router}} + \gamma\mathcal{L}_{\text{attn}}, \quad (12)$$

where $\alpha$, $\beta$ and $\gamma$ are the corresponding coefficient scores of each loss function.

# 4 Experiment

## 4.1 Experiment Setting

*4.1.1 Datasets.* We conduct experiments on a comprehensive benchmark, i.e., ALCE (Automatic LLM Citation Evaluation) benchmark [11]. It is designed to systematically evaluate LLMs' ability to generate text with accurate and contextually relevant citations. The three datasets in the ALCE benchmark are introduced as follows:

- **ASQA** [35] is designed for generative QA tasks with the goal to generate a single, concise answer summary from multiple supporting passages. It focuses on open-domain, multi-document QA and requires models to synthesize information, rather than simply extract text spans.
- **ELI5** [8] is a large-scale benchmark for long-form, open-domain question answering. It is designed to test a model's ability to generate detailed, explanatory answers to complex questions.
- **QAMPARI** [1] is designed to evaluate open-domain QA models on questions with multiple correct answers.

*4.1.2 Evaluation Metrics.* Following evaluations in the ALCE benchmark [11], given a question and retrieved documents, we evaluate the correctness and citation quality of responses.

- **Citation Quality**. We compute two key metrics: citation *Recall*, assessed via a natural language inference model called TRUE [13], to determine if citations sufficiently support statements and citation *Precision* (to check for irrelevant inclusions). We report their harmonic mean, the *Citation-F1* score, for a unified assessment.
- **Correctness**. It evaluates whether outputs accurately answer queries. *EM-recall* is used in the ASQA dataset and *Claims recall* is adopted to check sub-claim support in reference responses in the ELI5 dataset. For QAMPARI, *Recall-5* measures the correctness of questions with multiple correct answers.

*4.1.3 Compared Methods.* We selected three types of attribution LLM models for comparison, including:

- **P**rompt-based **I**n-**C**ontext-**L**earning (**P-ICL**): we compared three prompt-based methods proposed in the ALCE benchmark [11]. **VANILLA** simply provides top-5 documents and instructs the model to generate responses with citations. **SUMM** summarizes top-k input documents into a coherent overview via ChatGPT. **SNIPPET** extracts key snippets from the summarized top-k input documents.
- **I**terate-based **I**n-**C**ontext-**L**earning (**I-ICL**): Five representative I-ICL models are compared: **AnG** [34] uses three intuitive steps: content selection, sentence planning, and sequential sentence generation to ensure both conciseness and coverage, termed as "Attribute First, then Generate". **AAR** [22] employs LLMs to iteratively revise responses through a self-evaluation and refinement process. **Citation-Enhanced** [24] (shortened as C-Enhanced) generates answers and then retrieves support documents for each sentence, regenerating the answers iteratively until all sentences are supported by relevant documents. **Blueprint** [10] explores the attribution capabilities of plan-based models depending on how the plans are created. **Self-RAG** [2] leverages self-reflective retrieval capabilities, integrating retrieval, critique, and generation through self-evaluation.
- **F**ine-**T**uning-based models (**FT**): These methods primarily focus on constructing high-quality data for fine-tuning. **Front** [16] is a two-stage training framework that teaches LLMs to generate fine-grained grounded citations. **SynSciQA, SynSciQA+, and SynSciQA++** are three attribution datasets of varying quality constructed via a data synthesis pipeline proposed in [31] for LLM attribution fine-tuning, with 240,670 diverse natural language instructions [14] serving as the basic raw data.

Unlike existing fine-tuning models, which prioritize developing pipelines to synthesize or clean high-quality citation datasets for attribution training, our C$^2$-Cite takes a distinct approach. We emphasize designing effective citation learning methods that directly and accurately integrate citation sources into response generation, regardless of the dataset's preprocessing pipeline or synthesis method. Specifically, we fine-tune our model on the SynSciQA series datasets proposed in [31] for LLM attribution, leveraging their general-purpose design and minimal reliance on complex preprocessing pipelines.

*4.1.4 Implement Details.* We report the results of each method with its optimal hyperparameter settings on the validation data. For each baseline method, a grid search is applied to find the optimal settings, including the learning rate from {1e-5, 5e-5, 1e-4, 2e-4}. Our model C$^2$-Cite and SynSciQA are fine-tuned using LoRA (Low-Rank Adaptation) technique with rank 8. The parameters $\alpha$, $\beta$, $\gamma$ in the loss function (i.e., Eq. 12) are set to 0.2, 0.1 and 0.1, respectively. The results of In-context learning models and the Iterate-based models are implemented by the toolkit CiteLab [32]. The evaluation configurations are strictly aligned with those in the ALCE benchmark [11].

**Table 1: Experimental results on the ALCE benchmark on Llama3-8B. The underline represents the SOTA baseline Front according to the average Citation-F1 score on three datasets. Our model achieves the best citation F1 score, which is highlighted in bold. The improvement ratio is calculated by comparing C$^2$-Cite++ with the SOTA baseline Front model.**

| Type | Method | Datasets | | | | | | | | | | | | | |
|---|---|---|---|---|---|---|---|---|---|---|---|---|---|---|---|
| | | ASQA | | | | ELI5 | | | | QAMPARI | | | | Avg. F1 | Avg. Corr. |
| | | P(↑) | R (↑) | F1(↑) | Corr. (↑) | P(↑) | R (↑) | F1(↑) | Corr.(↑) | P(↑) | R (↑) | F1(↑) | Corr.(↑) | | |
| P-ICL | VANILLA | 60.0 | 55.9 | 57.9 | 38.6 | 33.2 | 35.5 | 34.3 | 9.2 | 12.4 | 16.9 | 14.3 | 18.6 | 35.3 | 22.1 |
| | SUMM | 69.0 | 67.2 | 68.1 | 23.5 | 40.5 | 36.4 | 38.4 | 9.4 | 27.4 | 30.5 | 28.9 | 14.0 | 45.1 | 15.6 |
| | SNIPPET | 68.9 | 66.7 | 67.8 | 20.2 | 48.4 | 48.5 | 48.4 | 6.8 | 28.4 | 30.3 | 29.3 | 7.7 | 48.5 | 11.6 |
| I-ICL | AnG | 40.7 | 43.8 | 42.2 | 18.4 | 9.1 | 10.6 | 9.8 | 10.8 | 17.9 | 19.5 | 18.7 | 3.6 | 23.6 | 10.9 |
| | AAR | 47.8 | 37.8 | 42.2 | 38.9 | 12.4 | 13.1 | 12.8 | 5.1 | 23.5 | 22.7 | 23.1 | 8.1 | 26.0 | 17.4 |
| | C-Enhanced | 40.8 | 30.9 | 35.2 | 31.0 | 30.2 | 23.7 | 26.6 | 10.7 | 37.3 | 21.1 | 26.9 | 4.2 | 29.6 | 15.3 |
| | Blueprint | 71.3 | 68.5 | 69.9 | 40.8 | 46.1 | 44.3 | 45.2 | 10.5 | 27.9 | 24.7 | 26.2 | 3.4 | 47.1 | 18.2 |
| | Self-RAG | 47.0 | 46.9 | 47.0 | 27.4 | 44.4 | 44.4 | 44.4 | 10.6 | 27.5 | 27.3 | 27.4 | 1.4 | 39.6 | 13.1 |
| FT | Front | 73.2 | 70.9 | 72.0 | 42.8 | 51.9 | 50.5 | 51.2 | 9.8 | 31.9 | 24.3 | 27.6 | 9.5 | 50.3 | 20.7 |
| | SynSciQA | 36.9 | 31.6 | 34.0 | 39.9 | 19.7 | 15.2 | 17.2 | 7.8 | 14.9 | 5.5 | 8.0 | 16.3 | 19.7 | 21.3 |
| | C$^2$-Cite | 72.8 | 65.2 | 68.8 | 45.2 | 52.4 | 43.3 | 47.4 | 10.0 | 26.8 | 14.3 | 18.5 | 16.6 | 44.8 | 23.9 |
| | SynSciQA+ | 43.7 | 41.0 | 42.3 | 40.6 | 21.5 | 18.8 | 20.1 | 8.2 | 15.1 | 13.5 | 14.3 | 15.9 | 25.6 | 21.6 |
| | C$^2$-Cite+ | 73.2 | 70.8 | 72.0 | 45.7 | 49.5 | 47.0 | 48.2 | 10.3 | 26.9 | 15.3 | 19.5 | 15.6 | 46.6 | 23.9 |
| | SynSciQA++ | 47.4 | 46.9 | 47.1 | 41.2 | 26.3 | 25.4 | 25.8 | 9.0 | 26.2 | 25.0 | 25.6 | 16.0 | 32.8 | 22.1 |
| | C$^2$-Cite++ | **76.3** | **74.4** | **75.3** | **48.8** | **55.8** | **53.9** | **54.8** | **10.7** | **32.8** | **26.9** | **29.6** | **13.8** | **53.2** | **24.3** |
| **Improvement↑** | | +4.2% | +4.9% | +4.6% | +14.0% | +7.5% | +6.7% | +7.0% | +9.2% | +2.8% | +10.7% | +7.2% | +45.3% | +5.8% | +17.4% |

## 4.2 Main Results

We make comprehensive evaluations of different attribution methods from two aspects: citation quality (i.e., Precision (**P**), Recall (**R**), **F1** score) and Correctness (**Corr.**). We report the experimental results on the Llama3-8B-Instruct model. The experimental results are shown in Table 1. We have the following observations:

(1) Prompt-based models (e.g., SUMM and SNIPPET) and iterative-based retrieval methods (e.g., Blueprint) both demonstrate strong performance, with the former standing out as optimal on the QAMPARI dataset. Enhanced by reinforcement learning, the fine-tuning based method Front achieves the best average performance on citation quality (i.e., 50.3%) among all compared baseline methods on three datasets.

(2) By incorporating contextual information, our model C$^2$-Cite achieves the best average performance on citation quality and response correctness over three datasets. Specifically, it outperforms the SOTA baseline Front by an average of 5.8% in citation quality and 17.4% in response correctness.

(3) Our framework is fine-tuned using the same data synthesis pipeline as SynSciQA. The improvements demonstrate that our framework maintains stable and superior performance across fine-tuning datasets of varying quality (SynSciQA, SynSciQA+, and SynSciQA++), fully embodying its advantage in balancing citation quality and response correctness.

(4) The optimization of the data synthesis pipeline is insufficient. Fine-tuning with carefully constructed synthetic attribution datasets (SynSciQA, SynSciQA+, SynSciQA++) may not guarantee optimal performance of attributed LLMs. In contrast, by incorporating the contextual-aware citation mechanism, our fine-tuning framework C$^2$-Cite achieves the SOTA performance on all compared methods in both citation quality and response correctness across three diverse datasets.

(5) Better citation performance does not always align with the higher response quality. There exists an inconsistency between these two metrics, which arises because citation markers disrupt pre-trained patterns, shifting attention to symbolic references over content coherence (This phenomenon is further analyzed in Section 5.3). For example, Front achieves high citation scores but exhibits lower response quality compared to methods (e.g., Blueprint in ELI5 dataset). Compared to SynSciQA+, C$^2$-Cite+ achieve better Citation-F1 while performing slightly worse in the response correctness metric on QAMPARI dataset.

# 5 Experimental Analysis

## 5.1 Efficiency Analysis

We evaluate model efficiency by Throughput (samples/second), which is defined as the number of samples processed per second. To ensure a comprehensive and fair comparison, we selected the best-performing model (i.e., SNIPPET, Blueprint, Front) in the three representative types of attribution LLM models. As illustrated in

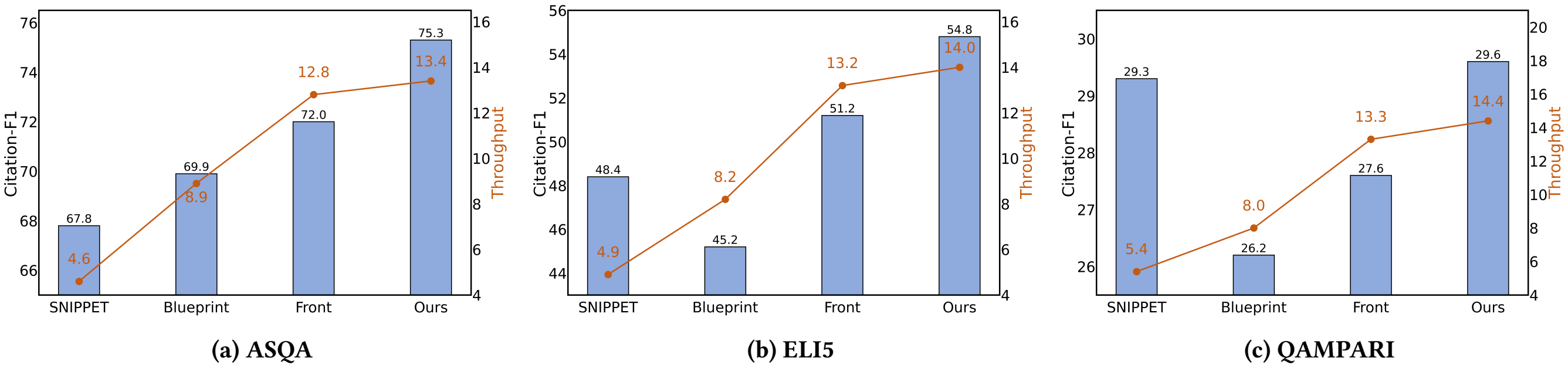

(a) ASQA (b) ELI5 (c) QAMPARI

**Figure 3: Efficiency comparison via Throughput (samples/second) on three datasets.**

Fig.3, which presents both the citation quality (measured by F1 score) and Throughput (samples/second) across compared models. Our framework c²-Cite++ achieves the highest Throughput, highlighting the unique advantages in achieving the SOTA effectiveness and superior operational efficiency in the LLM attribution task.

## 5.2 Ablation Study

To verify whether the contextual-aware embedding and contextual attentive augment are critical to the superior performance of our framework, we conduct ablation studies to validate the effectiveness of the two components. The experimental results of ablation studies are shown in Table 2.

- Contextual-Aware Embedding Incorporation (**w/o CAE**). We eliminate the contextual-aware embedding component by replacing the contextual input embeddings of citation markers with standard token embeddings while keeping all other experimental settings constant. We find that the removal of contextual-aware embedding results in a more notable degradation (i.e., 25.6% drop and 21.7 %) in Citation-F1 and Correctness, underscoring its greater importance in capturing contextual semantics to support accurate attribution and response generation.
- Contextual Attentive Augment (**w/o Attn**) . We disabled the attention mechanism by setting the attention weight parameter $\gamma = 0$ (see Eq. 12) while maintaining all other experimental configurations unchanged. It shows that removing the attention mechanism led to a 17% and 5.2% declines in Citation-F1 and response Correctness, with the drop being particularly pronounced on the QAMPARI dataset. This highlights the mechanism's critical role in enabling effective content generation with proper referencing attention.

## 5.3 Adverse Impact on Responses Correctness

Although providing attribution for generated content can offer a basis for the credibility of the generated content, we found that directly fine-tuning LLMs with attribution data containing citations may induce an adverse impact on the quality of the generated response content. As shown in Fig. 5, the response correctness in SynSciQA++ (i.e., Syn++) with citation attribution is lower than that without citation generation (i.e., Syn++ w/o Cite). This phenomenon may be attributed to the disruption of pre-trained generation patterns: introducing citation markers forces attention on symbolic references during response generation, disrupting semantic coherence or overemphasizing citation alignment over content fluency. In contrast, by incorporating contextual information of citation markers, our framework C²-Cite++ better supports answer generation, thereby maintaining high response generation correctness and enhancing citation quality.

**Table 2: Ablation Study: The Degradation (Degra.) of Citation-F1 score and Correctness by removing Contextual-Aware Embedding (CAE) and Contextual Attentive Augment (Attn) in our framework C²-Cite++.**

| Metrics | Datasets | Ours | w/o CAE | w/o Attn |
|---|---|---|---|---|
| Citation-F1 | ASQA | 75.3 | $48.8_{\mathbf{-35.2\%}\downarrow}$ | $70.4_{\mathbf{-6.5\%}\downarrow}$ |
| | ELI5 | 54.8 | $38.5_{\mathbf{-29.7\%}\downarrow}$ | $50.8_{\mathbf{-7.3\%}\downarrow}$ |
| | QAMPARI | 29.6 | $26.1_{\mathbf{-11.8\%}\downarrow}$ | $18.6_{\mathbf{-37.2\%}\downarrow}$ |
| | Avg.Degra. | - | 25.6% ↓ | 17% ↓ |
| Correctness | ASQA | 48.8 | $40.9_{\mathbf{-16.2\%}\downarrow}$ | $46.3_{\mathbf{-5.1\%}\downarrow}$ |
| | ELI5 | 10.7 | $9.6_{\mathbf{-10.3\%}\downarrow}$ | $10.3_{\mathbf{-3.8\%}\downarrow}$ |
| | QAMPARI | 13.8 | $8.5_{\mathbf{-38.5\%}\downarrow}$ | $12.9_{\mathbf{-6.6\%}\downarrow}$ |
| | Avg.Degra. | - | 21.7% ↓ | 5.2% ↓ |

## 5.4 Visualization of Contextual Incorporation

To further investigate how our citation markers, integrated with contextual information, impact content generation, we conducted a visual analysis of attention heatmaps for both native LLMs and our proposed model, with results shown in Fig. 6 (illustrating the last decoder's attention matrix of responses from different models). In the heatmap, qualitative attention values are visually represented: darker colors indicate higher attention weights, while lighter shades signify lower ones. Our framework C²-Cite demonstrates significantly stronger attention in the sentence $S_2$ following a citation marker toward the content of the preceding sentence $S_1$. For native LLMs-subfigure (a), Sentence $S_1$ and Sentence $S_2$ show minimal cross-attention. In contrast, our C²-Cite model-subfigure (b) exhibits substantially stronger attention from $S_2$ to $S_1$, highlighting how it fosters tighter inter-sentential connectivity via contextual citation incorporation. This disparity stems from our framework

**Query: Why do polar bears live at the Northern Pole and Penguins at the Southern?**

**Document 1: ...** All 17 species of penguins live in the Southern Hemisphere where there are no land predators. **...**

**Document 2: ...** For instance, both are aquatic species and are exceptional swimmers. ...

**Document 3: ...** Survival at the Poles Antarctic Survivor – the Emperor Penguin Polar bears and penguins never bump into each other! **...**

**Document 4: ...** Arctic Land Mammals It is vital for a mammal, being a 'warm-blooded' vertebrate, to keep warm in order to maintain its body at a constant temperature. **...**

**Document 5: ...**Are There Penguins In The North Pole? **...**

**Native Model**: Polar bears are confined to the Arctic due to specialized adaptations for hunting seals on sea ice **[1]**, while penguins thrive in the predator-free Antarctic environment where they evolved aquatic hunting skills **[5]**. Their distribution is mainly caused by evolutionary isolation over millions of years and the lack of land bridges between hemispheres.

**C2-Cite**: Polar bears live at the North Pole while penguins inhabit the South Pole primarily due to evolutionary adaptation to their respective environments. Penguins evolved in the Southern Hemisphere where they face no land predators **[1]**, developing swimming adaptations instead of flight capabilities. Their specialized hunting methods are optimized for Antarctic conditions **[3]**.

**Figure 4: A case study of attention pattens in Naive LLM vs. C$^2$-Cite++ in generated responses.**

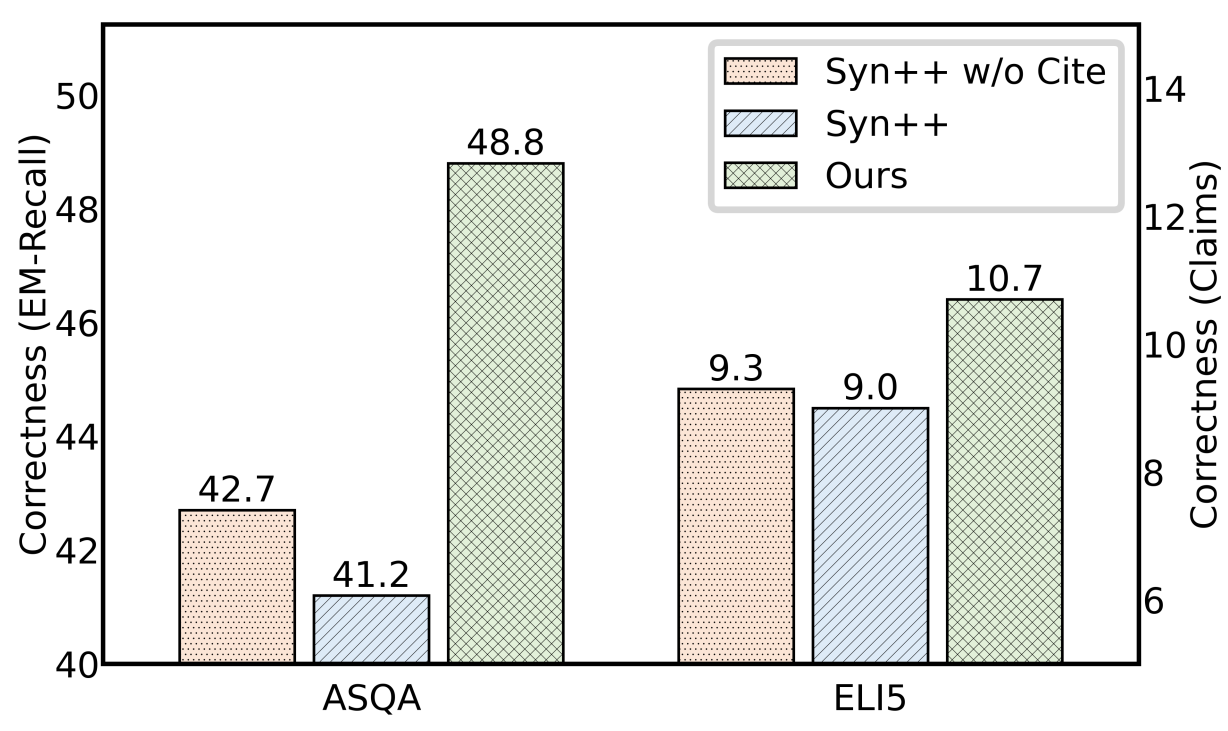

**Figure 5: Degradation of response correctness with citation-integrated training data. Our model enhances response quality by incorporating contextual information.**

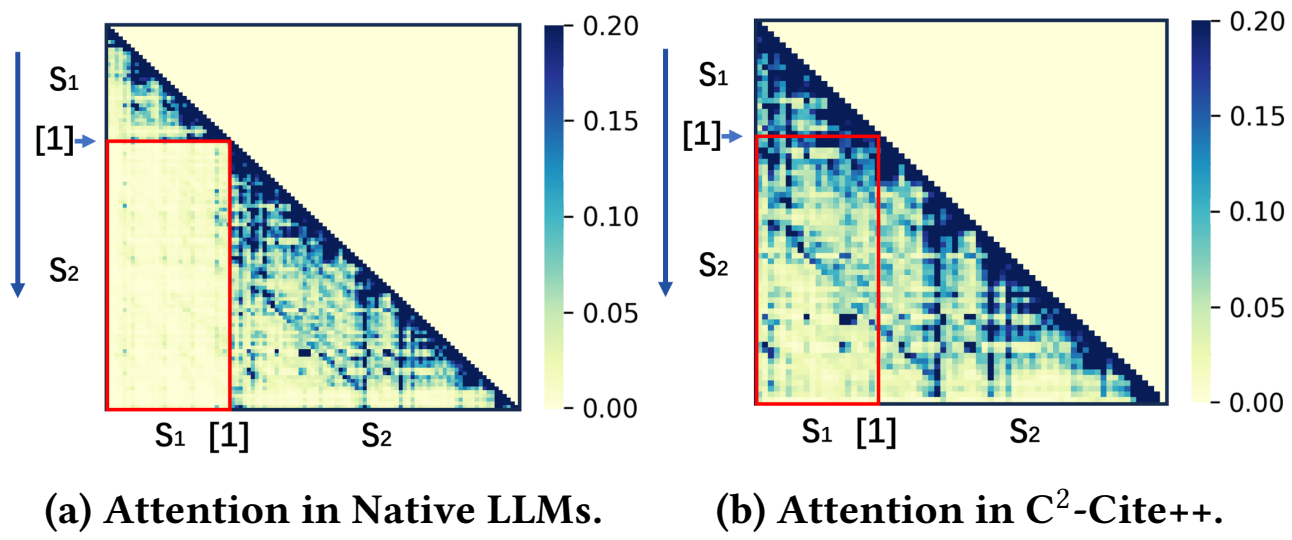

**(a) Attention in Native LLMs.** **(b) Attention in C$^2$-Cite++.**

**Figure 6: Attention heatmap of generated tokens in responses. By incorporating contextual information into the citation token between the two sentences, our model enables the generation of $S_2$ to pay more attention to $S_1$, ensuring semantic coherence in the response generation process.**

integrating contextual information from citation markers and conducting contextual attentive augmentation in the fine-tuning process, which enhances semantic coherence between segmented content and strengthens sequential attention mechanisms to maintain focus on prior generated content.

## 5.5 Case Study

To further demonstrate the influence of the incorporation of contextual information, Fig. 4 presents a case study of a specific query *"Why do polar bears live at the Northern Pole and Penguins at the Southern?"* along with the corresponding set of retrieved documents. In this visualization, words with high attention weights are indicated by the same color, reflecting strong semantic connections between them. In the response from the naive LLM, attention weights are primarily concentrated within individual sentences, while the sentences flanking the citation symbols (e.g., [1], [5]) appear relatively isolated. This weak inter-sentential connection disrupts the overall coherence of the response, as the shifts between cited and non-cited content lack sufficient semantic continuity.

In contrast, in the response generated by our C$^2$-Cite++ framework, the sentences before and after the citation markers (e.g., [1]) exhibit significantly higher relevance. This is evident from the concentrated attention weights between these adjacent sentences, ensuring that the semantic flow remains coherent even when citations are inserted. For instance, the statement about penguins "developing swimming adaptations instead of flight capabilities [1]" is tightly linked to the preceding context about their evolutionary environment in the Southern Hemisphere, creating a smooth logical progression. This comparison shows that C$^2$-Cite++ links responses to sources by strengthening contextual bonds between citation-related sentences, enhancing the quality of the generated responses.

## 6 Conclusion

Our work overcomes a critical limitation in current attributed LLMs: they treat citation symbols as generic placeholders, failing to capture contextual semantics. This oversight results in disjointed references, poor knowledge integration, and reduced response quality. We introduce $C^2$-Cite, a novel framework that addresses poor citation integration in LLMs by explicitly modeling contextual semantic links between citation markers and their referenced content. Our model encodes document contextual content into citation symbols, transforming passive placeholders into active knowledge pointers via contextual alignment, ensuring seamless citation integration and enhancing response correctness and attribution quality.

## 7 Ethical Considerations

This study utilized publicly available datasets (e.g., ASQA, ELI5, QAMPARI, SynSciQA) in compliance with their usage terms, and the proposed technical framework, focused on improving capabilities and efficiency in attribution tasks, is designed without functionalities that could pose ethical risks, though we acknowledge the need for context-specific safeguards in real-world deployment.

## Acknowledgements

This work was supported by the National Key Research and Development Program of China (No. 2023YFC3303800).